\documentclass[%
reprint,
 amsmath,amssymb,
aps,
]{revtex4-2}

\usepackage{graphicx}
\usepackage{dcolumn}
\usepackage{bm}

\usepackage{enumitem} 
\usepackage{amsmath}
\usepackage{color}
\usepackage{soul}

\def\e{\begin{equation}}
\def\f{\end{equation}}
\def\_#1{{\bf #1}}

\def\.{\cdot}

\def\r#1{(\ref{eq:#1})}
\def\@#1{_{\rm #1}}

\begin{document}


\title{%
Coherent Asymmetric Absorbers}

\author{F.S. Cuesta$^{1}$}%
 \email{francisco.cuestasoto@aalto.fi}
\author{A.D. Kuznetsov$^{1,2}$}
\author{G.A. Ptitcyn$^{1}$}%
\author{X. Wang$^{1}$}%
\author{S.A. Tretyakov$^{1}$}%
\affiliation{%
 $^{1}$Department of Electronics and Nanoengineering, Aalto University, P.O. Box 15500, Aalto FI-00076, Finland \\
 $^{2}$School of Electronic Engineering, HSE University, Russia
}%

\date{\today}

\begin{abstract}
Most  applications  of  metasurfaces  require  excitation and control of both electric  and  magnetic  surface  currents. For such purpose,  the metasurface must have a finite  thickness to handle magnetic surface currents.  
For  metasurface sheets of negligible thickness that offer only electric response, coherent illumination can compensate the need to create discontinuities of the tangential electric field component using magnetic surface currents. Most of known coherent metasurfaces are space-uniform and can control only plane-wave absorption and specular reflection.  However, it is also known that periodical space-modulated (inhomogeneous) metasurfaces can be used to realize anomalous reflection, refraction, and other useful effects. In this paper, we propose the concept of a coherently-illuminated space-modulated metasurface that functions as a coherent asymmetric absorber. We study its behaviour under non-ideal illuminations and suggest  applications related with sensing.
\end{abstract}

\maketitle


\section{Introduction}

The first step in designing  metasurface devices is to determine the surface impedance that corresponds to the desired scattering for a predefined illumination (e.g., \cite{surface_EM,simovski_tretyakov_2020}). For different design targets, the required properties of the metasurface can be different, requiring chirality (for polarization conversion, for example), time modulation (for frequency conversion or nonreciprocal response), space modulation (for engineering reflection and refraction directions), or other  properties. In the case of periodical space modulations, the availability and propagation directions of the non-specular modes are determined by the period of the surface impedance variations, as described by the Floquet theory \cite{Ishimaru_Floquet}. However, the power distribution between specular and available non-specular modes is determined by the surface impedance variation profile  \cite{Wang_2020_Channels}. Because of that, space-modulated metasurfaces can produce such effects as anomalous reflection and refraction \cite{Li_2014,Tang2014,Diaz2017_generalizedreflection,Sell2018,simovski_tretyakov_2020}, retroreflection \cite{Jia_2017,Asadchy2017,Li2018}, or asymmetric absorption \cite{Wang_2018_Asymmetry}.

At the conceptual level, a metasurface behaves as a sheet creating discontinuities of fields or as an impenetrable boundary. If the objective is to control reflection only, metasurfaces are often realized as an impedance sheet above a metallic ground plane, realizing an effective boundary at the plane of the impedance sheet \cite{Tretyakov2003}. On the other hand, metasurfaces designed for control of transmission do not have a ground plane, behaving as sheets carrying electric and magnetic surface currents \cite{simovski_tretyakov_2020}. Many functionalities cannot be realized using sheets supporting only electric surface currents (for microwave applications, usually realized as single arrays of thin metal patches), for example, perfect absorption or retroreflection. In these cases, metasurfaces must carry also magnetic surface currents, which requires the use of at least two parallel arrays or some volumetric particles \cite{Asadchy_Broadband_Metasheets}. However, coherent metasurfaces can produce these desired effects using a secondary illumination source, even if the metasurface sheet is negligibly thin and no magnetic current can be excited  \cite{Lv_2018,Feng_2018,Zou_2019,Baranov_2017,Nie_2014,Cuesta_2021_coherent_retroreflector}.
Space-modulated metasurfaces boundaries, realized as finite-size layers carrying both electric and magnetic surface currents, can behave as asymmetric absorbers \cite{Wang_2018_Asymmetry,Song2019,Wang2019,Dong2020}. 

\begin{figure}[t]
\centering
\includegraphics[width=1\linewidth]{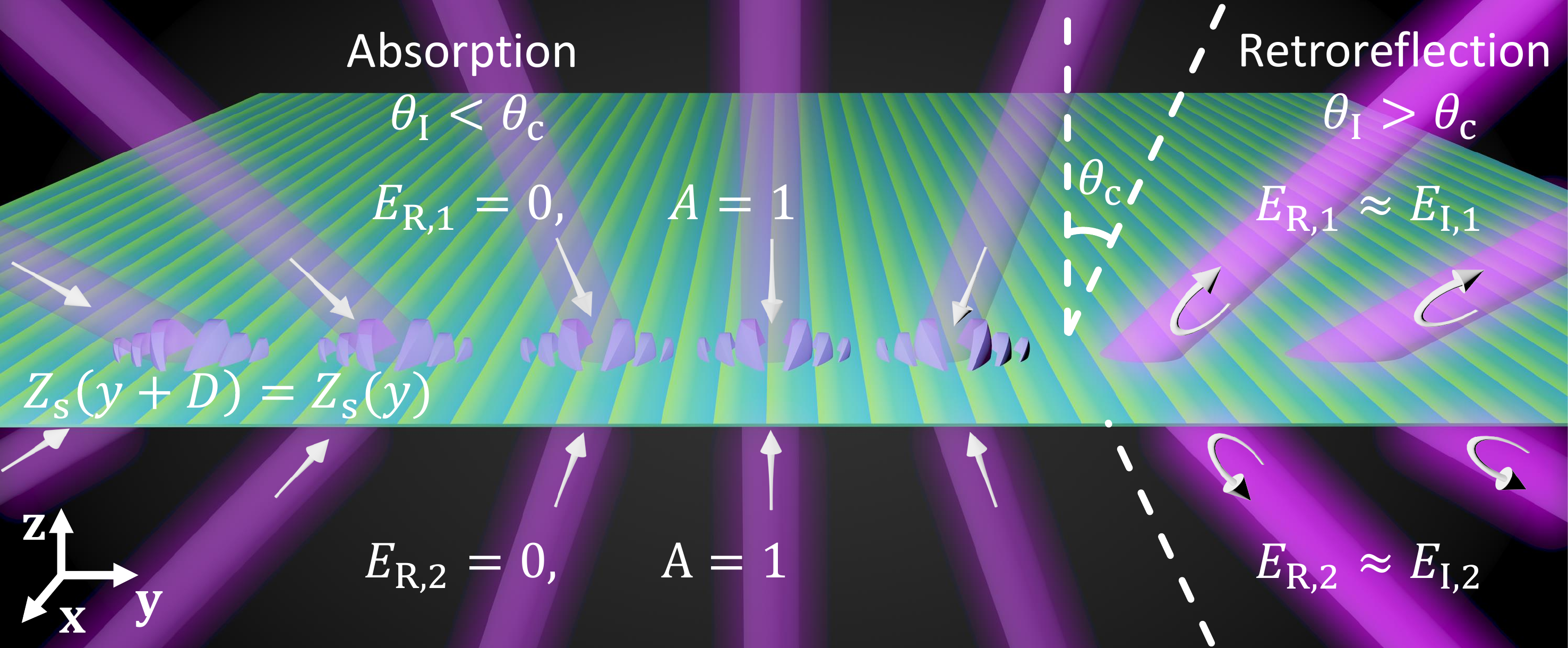}
\caption{Concept of a coherent asymmetric absorber, which for different angles of incidence behaves as an absorber or as a reflective surface. 
The cutoff angle $\theta\@{c}$ is determined by the metasurface period. \label{fig:coherent_concept} }
\end{figure}

In this paper, we introduce the concept of a space-modulated, single-sheet coherent asymmetric absorber: a device that behaves as an absorber or as a retroreflector depending on the angle of incidence of illumination by two coherent waves, as illustrated in Fig.~\ref{fig:coherent_concept}. This work focuses on describing how the angle of incidence and variations in the magnitudes and phases of the coherent sources affect the response of the metasurface. In addition, this paper proposes applications where ideal and non-ideal coherent illuminations can be exploited for sensing and power absorption.

\section{Coherent retroreflection and absorption in space-modulated sheets}

\begin{figure}[h]
\centering
\includegraphics[width=0.95\linewidth]{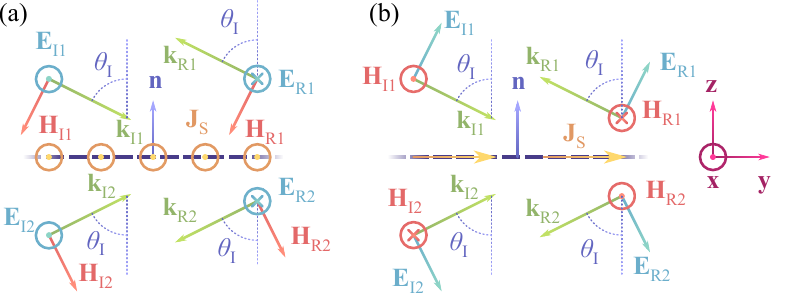}
\caption{\label{fig:reflect_modes} Incident and reflected plane waves from coherent retroreflective electromagnetic metasurfaces which are designed for (a)--transverse electric and (b)--transverse magnetic polarizations. }
\end{figure}

Let us consider a flat negligibly-thin metasurface that divides the space into two regions. The metasurface is illuminated by two plane waves at the same angle of incidence $\theta\@{I}$, but incident from each region separately. We start the analysis by designing a metasurface impedance profile that ensures perfect retroreflection of both illuminating waves at the design angle $\theta\@{I}=\theta\@{ms}$, see illustration in Fig.~\ref{fig:reflect_modes}. The incident waves can be polarized either as transverse electric (TE) or transverse magnetic (TM) waves. This paper will focus on TE-illumination scenarios, as shown in Fig.~\ref{fig:reflect_modes}(a), but the same principles can be also applied to  the TM-case, illustrated in  Fig.~\ref{fig:reflect_modes}(b). The fields for a TE-polarized illumination read
\begin{equation}
\begin{split}
&\_E\@{I,1}=E\@{I,1} e^{-j k_0 \left[y {\rm sin} (\theta\@{I})-z {\rm cos} (\theta\@{I})\right]} \_x,\cr
&\_H\@{I,1}=-\dfrac{E\@{I,1}}{\eta_0}\Big[{\rm cos} (\theta\@{I})\_y+{\rm sin} (\theta\@{I})\_z\Big]e^{-j k_0 \left[y {\rm sin} (\theta\@{I})-z {\rm cos} (\theta\@{I})\right]},
\end{split}
\label{eq:TE_incw_1}
\end{equation}
\begin{equation}
\begin{split}
&\_E\@{R,1}=-E\@{R,1} e^{j k_0 \left[y {\rm sin} (\theta\@{I})-z {\rm cos} (\theta\@{I})\right]} \_x,\cr
&\_H\@{R,1}=-\dfrac{E\@{R,1}}{\eta_0}\Big[{\rm cos} (\theta\@{I})\_y+{\rm sin} (\theta\@{I})\_z\Big]e^{j k_0 \left[y {\rm sin} (\theta\@{I})-z {\rm cos} (\theta\@{I})\right]},
\end{split}
\label{eq:TE_reflc_1}
\end{equation}
in region 1, and 
\begin{equation}
\begin{split}
&\_E\@{I,2}=E\@{I,2} e^{-j k_0 \left[y {\rm sin} (\theta\@{I})+z {\rm cos} (\theta\@{I})\right]} \_x,\cr
&\_H\@{I,2}=\dfrac{E\@{I,2}}{\eta_0}\Big[{\rm cos} (\theta\@{I})\_y-{\rm sin} (\theta\@{I})\_z\Big]e^{-j k_0 \left[y {\rm sin} (\theta\@{I})+z {\rm cos} (\theta\@{I})\right]},
\end{split}
\label{eq:TE_incw_2}
\end{equation}
\begin{equation}
\begin{split}
&\_E\@{R,2}=-E\@{R,2} e^{j k_0 \left[y {\rm sin} (\theta\@{I})+z {\rm cos} (\theta\@{I})\right]} \_x,\cr
&\_H\@{R,2}=\dfrac{E\@{R,2}}{\eta_0}\Big[{\rm cos} (\theta\@{I})\_y-{\rm sin} (\theta\@{I})\_z\Big]e^{j k_0 \left[y {\rm sin} (\theta\@{I})+z {\rm cos} (\theta\@{I})\right]},
\end{split}
\label{eq:TE_reflc_2}
\end{equation}
in region 2, where $E\@{I}$ and $E\@{R}$ are the amplitudes of the incident and reflected waves, respectively; $k_0$ is the wavenumber of the incident waves and $\eta_0$ is the characteristic impedance of the surrounding medium. The metasurface must be tailored such that only a single non-specular mode gets excited when the metasurface is illuminated at the design angle $\theta\@{ms}$. The resulting reflected mode propagates back in the same direction as the incident wave,  with a predetermined relative reflection amplitude $\Gamma$ and phase $\phi\@R$  with respect to the incident waves ($E\@{R1}=E\@{I1} \Gamma \exp \left[ j\phi\@{R} \right]$, $E\@{R2}=E\@{I2} \Gamma \exp \left[ j\phi\@{R} \right]$). The excess of energy supplied by the sources is absorbed by the metasurface. 

The metasurface can be modeled as a sheet of electric surface current density $\_J\@{s}=\_E/Z\@{s}$, where $\_E$ is the tangential component of the macroscopic electric field at the metasurface plane, and  $Z\@{s}$ is the surface impedance. Since no magnetic surface currents are induced, the metasurface does not produce any discontinuity in the tangential electric field. In this case, symmetric perfect retroreflection can be realized only if both incident waves have the same amplitude and phase ($E\@{I1}=E\@{I2}$), so that also the reflected waves have equal complex amplitudes ($E\@{R1}=E\@{R2}$). Therefore, for a given incidence direction, the surface impedance required for perfect coherent retrorefleciton can be determined by satisfying the boundary conditions for the total tangential electric and magnetic fields (the electric surface current density equals to the jump of the tangential magnetic field). For TE or TM illumination, the surface impedance profile reads, respectively, 
\begin{subequations}
\label{eq:zs}
\begin{equation}
  Z_s^{\rm TE} \left(y\right) = \frac{\eta_0}{2 \cos \theta\@{ms}} \frac{\exp\left[-2 j k_0 y \sin \theta\@{ms}\right]-\Gamma \exp\left[j \phi_R\right]}{\exp\left[-2 j k_0 y \sin \theta\@{ms}\right]+\Gamma \exp\left[j \phi_R\right]},\label{eq:zs_te}
\end{equation}    
\begin{equation}
  Z_s^{\rm TM} \left(y\right) = \frac{\eta_0 \cos \theta\@{ms}}{2} \frac{\exp\left[-2 j k_0 y \sin \theta\@{ms}\right]+\Gamma \exp\left[j \phi_R\right]}{\exp\left[-2 j k_0 y \sin \theta\@{ms}\right]-\Gamma \exp\left[j \phi_R\right]}.\label{eq:zs_tm}
\end{equation}
\end{subequations}

Through inspection, it can be found that the impedance profile is a periodic function with the period $D=\lambda_0 /\left(2 \sin \theta\@{ms}\right)$, where $\lambda_0$ is the reference design wavelength. A change of the reflection phase $\phi_R$ corresponds to  a lateral shift of the impedance position. 

\begin{figure*}[t]
\centering \includegraphics[width=0.8\linewidth]{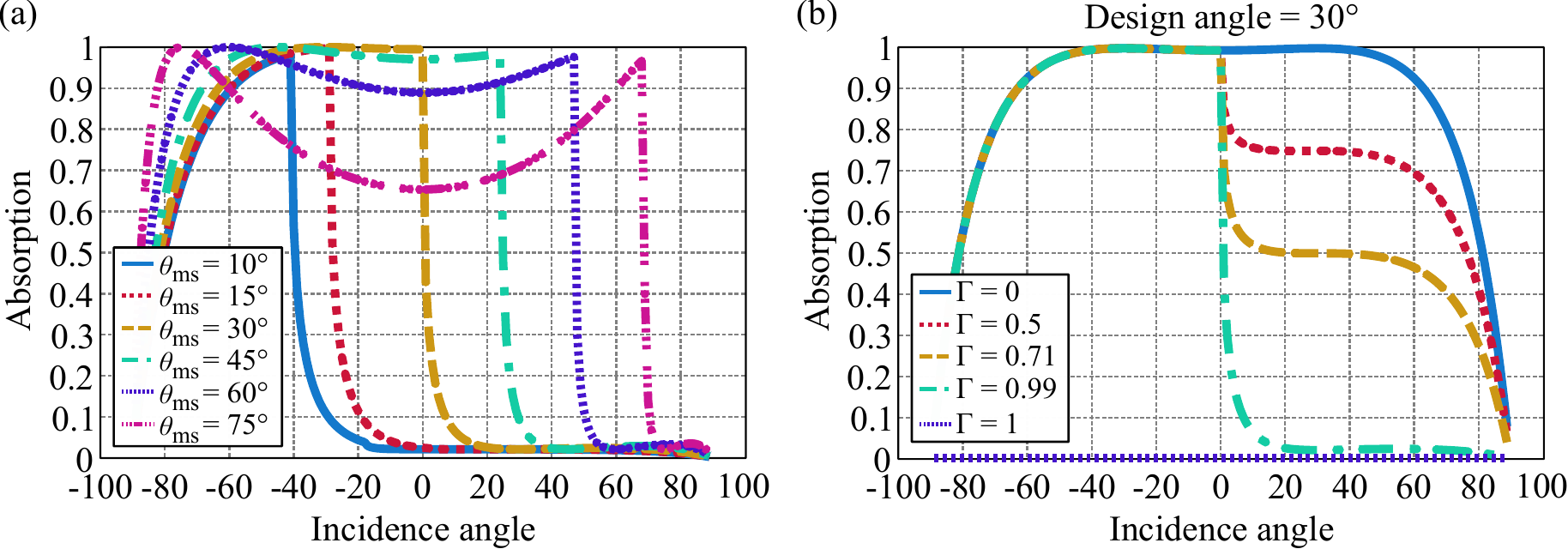}
\caption{(a) A metasurface designed to create a non-specular mode, acting as a retroreflector at the design  incidence angle $\theta\@{ms}$, has an asymmetric behavior in the form of an absorption region, where non-specular modes only exist as  evanescent waves. This example considers surface impedances corresponding to $\Gamma=0.99$. (b) In the angular region where the non-specular mode can propagate, its amplitude can be determined by  the reflection coefficient $\Gamma$ as a design parameter. The structure becomes lossless and symmetric if the value of $\left\vert\Gamma\right\vert=1$ is used, however, even for low-loss sheets with the reflection coefficients close to this value there is strong contrast between absorption and reflection regions.} \label{fig:scattering_results}
\end{figure*}

Numerical results of Fig.~\ref{fig:scattering_results}(a) for coherent retroreflectors with different design angles  $\theta\@{ms}$ reveal that the angular dependence of absorption is strongly asymmetric.  The absorption is high in the angular region where there is only one propagating mode available (specular reflection), and it is weak where the non-specular mode is excited as a surface  wave. By changing the design angle of incidence  $\theta\@{ms}$, it is possible to control these angular regions, delimited by the cutoff angle $\theta\@{c}$. The high-absorption regime mostly corresponds to illuminations with ``negative'' angles of incidence (tilted to the left from the normal to the metasurface plane, as defined in Fig.~\ref{fig:reflect_modes}). 

This asymmetric behaviour allows us to design metasurfaces that exhibit strong coherent retroreflection for positive angles of incidence but strong coherent absorption for negative ones, as is illustrated in Fig.~\ref{fig:scattering_results}(a). An additional design feature is  a possibility of controlling the reflection coefficient of the non-specular mode without disturbing the asymmetric absorption, as portrayed in Fig.~\ref{fig:scattering_results}(b). The asymmetry is present for almost all values of the amplitude of the design reflection coefficient $\Gamma$, except for the extreme values $\left\vert\Gamma\right\vert=1$ and $\Gamma=0$, which correspond to already known cases of the perfect coherent retroreflector \cite{Cuesta_2021_coherent_retroreflector} and the perfect coherent absorber \cite{Fan2014}, respectively. This asymmetric angular dependence in absorption is similar to the lossy retroreflective boundaries of Ref.~\cite{Wang_2018_Asymmetry}, illuminated by only one wave.

\section{Asymmetric Illumination}

\begin{figure}[t]
\centering
\includegraphics[width=0.65\linewidth]{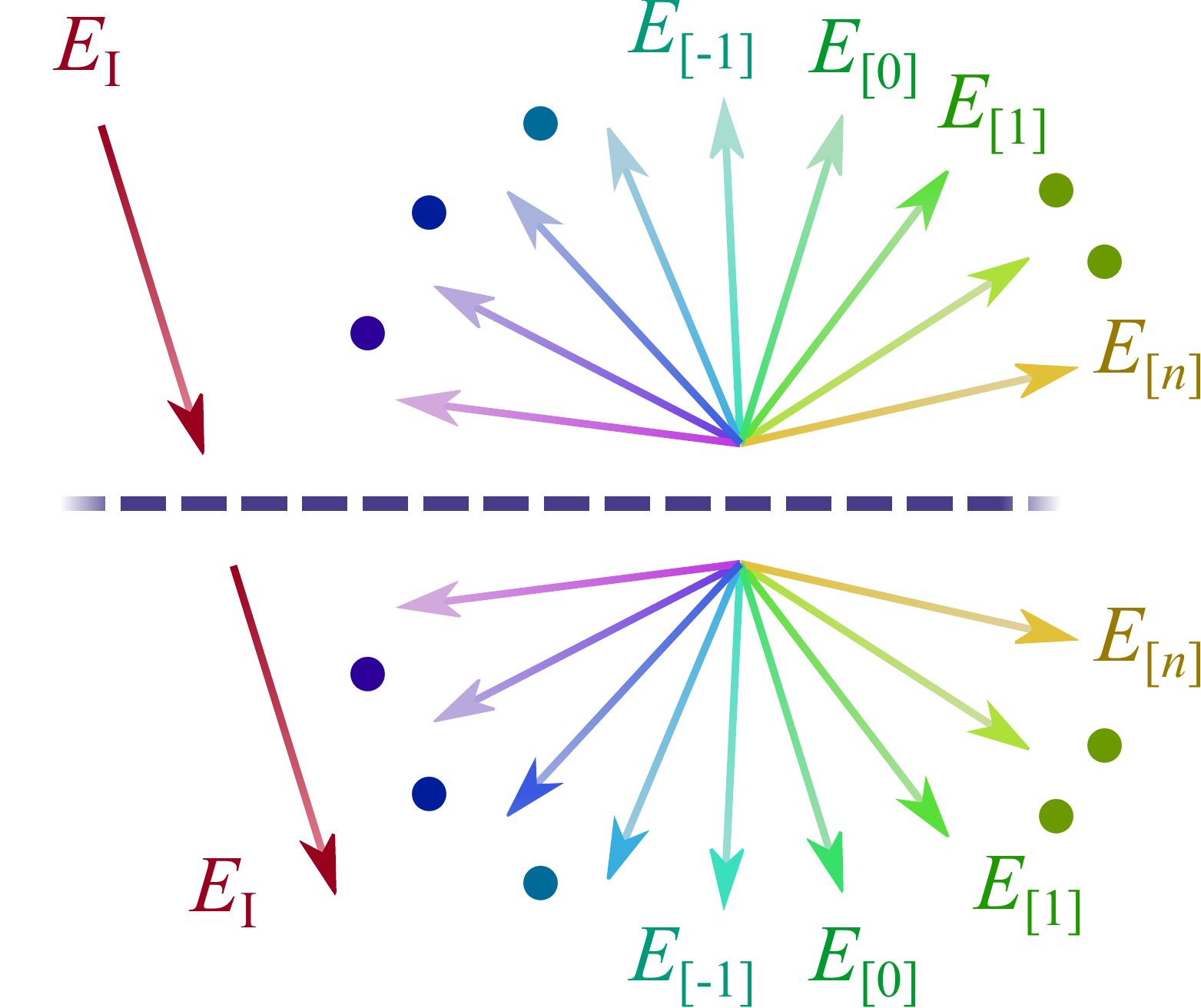}
\caption{\label{fig:multiscatt_modes} As described by the Floquet theory, uniform illumination of a periodic surface excites a certain range of specular and non-specular waves. While the mode range is defined by the period of the surface and the incidence angle of the exciting wave, the amplitude of each excited mode is determined by the surface impedance profile.}
\end{figure}

The most attractive feature  of coherent metasurfaces is a possibility to tune the response by varying phase or amplitude differences between the illuminating waves. To this end, here we analyse the response of coherent retroreflectors 
illuminated by an arbitrary set of two plane waves. When a sheet designed to function as a coherent retroreflector for a given incidence angle is illuminated by any other source, the reflected field contains infinitely many plane-wave components, as portrayed in Fig.~\ref{fig:multiscatt_modes}. Let us consider a periodical metasurface with the surface impedance $Z_s$  [in particular, given by Eq.~\eqref{eq:zs_te} or \eqref{eq:zs_tm}] with the period $D$, illuminated by a single TE-polarized wave (similar conclusions are achieved for the TM-polarization) with the angle of incidence $\theta\@{I}$. As described by the Floquet theory, the field on the metasurface plane can be considered as a combination of propagating and evanescent plane-wave modes: 
\begin{equation}
E=E\@{I} e^{-jy k\sin \theta_{i}} + \sum_{n=-\infty}^{\infty} E_{[n]} e^{-jy k_{y[n]}},
\end{equation}
where $E_{[n]}$ is the amplitude of the $n$-th Floquet harmonic of the field expansion, defined by the wavevector components $k_{y[n]}=k \sin \theta_{i}+ 2 \pi n / D$ and $k_{z[n]}=\sqrt{k^2-k_{y[n]}^2}$. For convenience, we define the mode angle $\theta_{[n]}=\arcsin{\left(k_{y[n]}/k\right)}$. Note that this angle is imaginary for evanescent components, and becomes real at the cutoff angle $\theta\@{c}$ where $k_{y[n]}= \left\vert k \right\vert$. The tangential electric field is continuous across the sheet, and the jump of the tangential magnetic field equals to the surface current density of the induced electric current: 
\begin{equation}
\label{eq:gen_bound_cond}
  \sum_{n=-\infty}^{\infty} 2E_{[n]} \cos \theta_{[n]} e^{-jy k_{y[n]}} = -\eta_0 J_{s}.
\end{equation}
\begin{table*}
\begin{equation}
\begin{split}
    \sum_{n=-\infty}^{\infty} \left(2 \cos \theta_{[n],1}+\dfrac{\eta_0}{Z_s} \right) E_{ [n],1} e^{-jy k_{y[n],1}} & + \sum_{n=-\infty}^{\infty}  \left(2 \cos \theta_{[n],2}+\dfrac{\eta_0}{Z_s} \right) E_{ [n],2} e^{-jy k_{y[n],2}}\\
    & = -\dfrac{\eta_0}{ Z_s} \left[ E\@{I,1} e^{-jy k\sin \theta_{i,1}}+ E\@{I,2} e^{-jy k\sin \theta_{i,2}}\right]\label{eq:coherent_scattering_relation}
\end{split}    
\end{equation}
\end{table*}

Because the sheet response is linear, the current excited by two coherent plane waves can be found as the sum of two independent Floquet expansions corresponding to the two different periods fixed by the two different incidence angles. The boundary condition $Z_s\_J_e=\_E  $ takes the general form of  Eq.~\r{coherent_scattering_relation}. In the special case when the two incidence angles are the same, the periods of the two expansions are equal, and the two series can be combined into one. For  this scenario, Eq.~\r{coherent_scattering_relation} can be rewritten by combining pairs of  scattered harmonics ($E_{ [n]}=E_{ [n],1}+E_{ [n],2}$), and the boundary condition takes the form
\begin{equation}
\begin{split}
\sum_{n=-\infty}^{\infty} \left(2 \cos \theta_{[n]}+\dfrac{\eta_0}{Z_s} \right) & E_{[n]} e^{-jy k_{y[n]}}\\
&= -\dfrac{\eta_0}{Z_s} e^{-jy k\sin \theta_{i}}\left[ E\@{I1}+ E\@{I2}\right],\label{eq:coherent_scattering_relation_mirrored}
\end{split}
\end{equation}
where each scattering mode does not only depend on the interaction between the fields and the metasurface, but also on the direct interaction between the incident waves at the boundary. 

In order to understand the effect of phase amplitude differences between the two illuminaitng waves, the amplitude of the source in region~2 can be defined as proportional to the one in region~1: $E\@{I2}= E\@{I1}\rho\@{I2} \exp\left[j \phi\@{I2}\right]$, where $\rho\@{I2} $ is the amplitude ratio. The total incident field at the metasurface plane is then written as 
\begin{equation}
    E\@{tot}=E\@{I1} \left[1+\rho\@{I2} e^{j \phi\@{I2}}\right].\label{eq:coherence_function}
\end{equation}
The metasurface is functioning as a perfect coherent retroreflector when the two waves have equal amplitudes ($\rho\@{I2} = 1$) and phases ($\phi\@{I2}=0$). Otherwise, the metasurface is ``mismatched'' with respect to the sources, and its response can be dramatically different. In the limiting case of the equal amplitude but opposite phase, that is, when $\rho\@{I2} = 1$ and $\phi\@{I2}=\pi$, there is no interaction between the sheet and the incident waves.  

\section{Numerical Characterization}
\label{secnum}

The exact values of the scattered waves amplitudes can be determined for lossless metasurfaces by solving Eq.~\eqref{eq:coherent_scattering_relation} in combination with the power conservation law for the propagating modes ($\left\vert k_{y[n]} \right\vert \leq k$). On the other hand, if the metasurface is lossy or it has an active behavior, power conservation is not usable as the total scattered power is not equal to the power radiated by the metasurface. For these scenarios, approximate solutions can be found using numerical means based on the Fourier series expansion of the surface impedance $Z_s$ in combination with the mode-matching method developed in Refs.~\cite{Wang2020_design,Wang_2020_Channels}. The absorption in the  metasurface is found by subtracting the power reflected into free space from the total incident power.

\begin{figure*}[t]
\centering \includegraphics[width=0.8\linewidth]{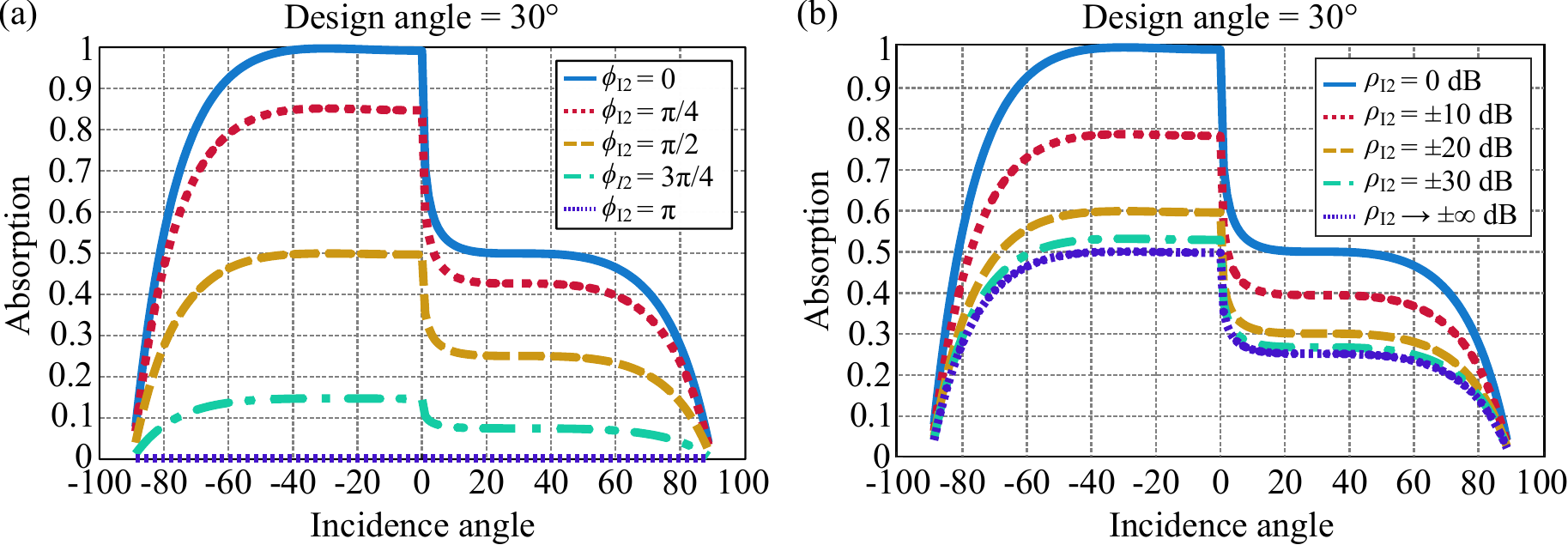}
\caption{(a) A coherently-illuminated metasurface designed with equivalent electric current densities in mind is matched with the  sources when both waves illuminate the surface at the same phase. Any variation in the phase will create a mismatch, leading to some specular reflection. (b) Similarly, the sources are matched with the metasurface as long as both plane waves illuminate the surface with the same amplitude. Any difference in the amplitude breaks the system symmetry, exciting undesired specular reflection. Examples correspond to surface impedances with $\Gamma=0.71$ and $\theta\@{ms}=30^{\circ}$. } \label{fig:coherence_scattering_results}
\end{figure*}

Results of Fig.~\ref{fig:scattering_results} show the behaviour of the proposed structure with a combination of matching sources, as described in Eq.~\r{coherence_function}. The strong absorption regime for ``negative angles'' ($\theta\@{I}<0$) is achieved by preventing the excitation of the first available non-specular mode ($n=-1$), leaving only specular reflection ($n=0$). On the other hand, Figs.~\ref{fig:coherence_scattering_results}~(a) and (b) show how a mismatch between the two coherent sources affect absorption in the metasurface when different phase shifts $\phi\@{I2}$ and amplitude ratios $\rho\@{I2}$  are considered, respectively. Obviously, the metasurface behaves as a lossless, fully transparent structure when the amplitudes are equal and the phase difference is equal to $\pi$. When one of the two amplitudes is much larger than the other, we approach the regime of a single-wave excitation of a sheet of negligible thickness, in which case absorption cannot be higher than  $50\%$ of the incident power.
 
\section{Potential applications of asymmetric coherent absorbers}

\begin{figure}[tb]
\centering \includegraphics[width=0.9\linewidth]{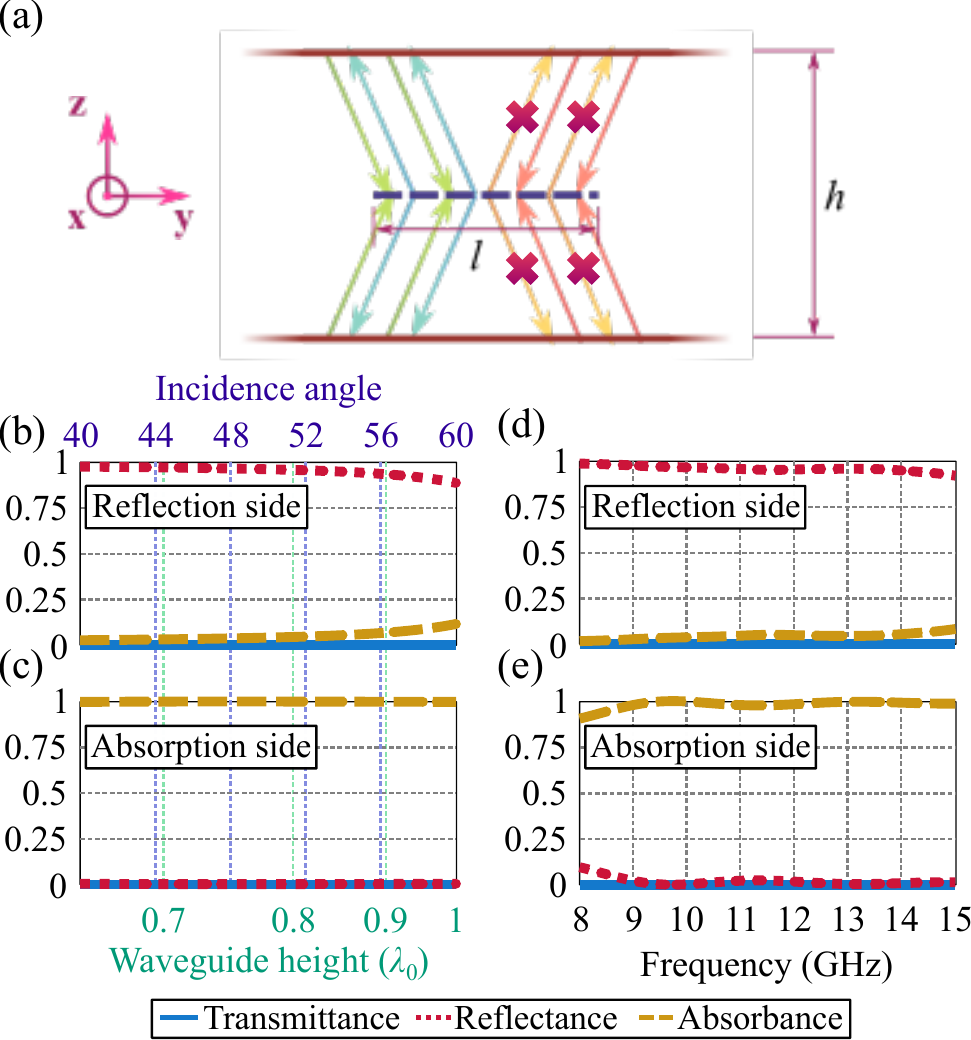}
\caption{(a) By inserting a coherent asymmetric absorber inside a parallel-plate waveguide, it is expected to transfer its properties into the waveguide for TE$_1$ mode. (b) and (d) If the metasurface is designed for $\theta\@{ms}=30^{\circ}$, one end of the waveguide becomes highly reflective. (c) and (e) On the other hand, the waveguide absorbs the TE$_1$ wave when it is fed from the opposite end of the waveguide. For this example, the metasurface was designed with $\theta\@{ms}=30^{\circ}$, $\Gamma=0.99$, for a reference frequency of $10$ GHz and with total length of $l=10D$.} \label{fig:asymmetric_waveguide}
\end{figure}

Asymmetric absorption and dramatic differences of response of  asymmetric coherent absorbers under coherent and incoherent illuminations can be exploited in various applications. First, let us consider the setup of Fig.~\ref{fig:asymmetric_waveguide}(a), where a sample of such metasurface is placed in the symmetry plane of a parallel-plate waveguide, supporting the TE$_1$ mode. The field of this mode can be decomposed into two coherent plane waves \cite{Pozar_2004}. By changing the separation between the plates $h$, the angle of incidence of these partial waves on the metasurface can be tuned.  In the considered example, the metasurface is designed with $\theta\@{ms}=30^{\circ}$ ($\theta\@{c}=0^{\circ}$) and $\Gamma=0.99$, resulting in high reflection for positive angles and strong absorption for the opposite tilt, as shown in Fig.~\ref{fig:scattering_results}(a). In this configuration, the waveguide inherits the properties of the metasurface as a retroreflector or as an absorber, depending on which end of the waveguide is excited, as portrayed in Figs.~\ref{fig:asymmetric_waveguide}(b)-(c). This behavior is stable over a wide frequency range where higher-order modes beyond TE$_1$ cannot propagate, as is seen in Fig.~\ref{fig:asymmetric_waveguide}(d)--(e). However, close to the cutoff frequency of the TE$_1$ mode, the response is affected by the difference between the waveguide characteristic impedance with and without the metasurface, resulting in high reflection from both sides. Let us note that this setup does not break reciprocity, and transmission through the device is symmetric. 

We see that this configuration functions as a compact and broadband microwave absorber in a waveguide that does not contain any terminating boundary and can freely pass high-frequency electromagnetic radiation or, for example, flow of air or water. Also, for waves propagation into the opposite direction, the device works as broadband reflector, although there is no reflecting wall.

\begin{figure*}[tb]
\centering \includegraphics[width=1\textwidth]{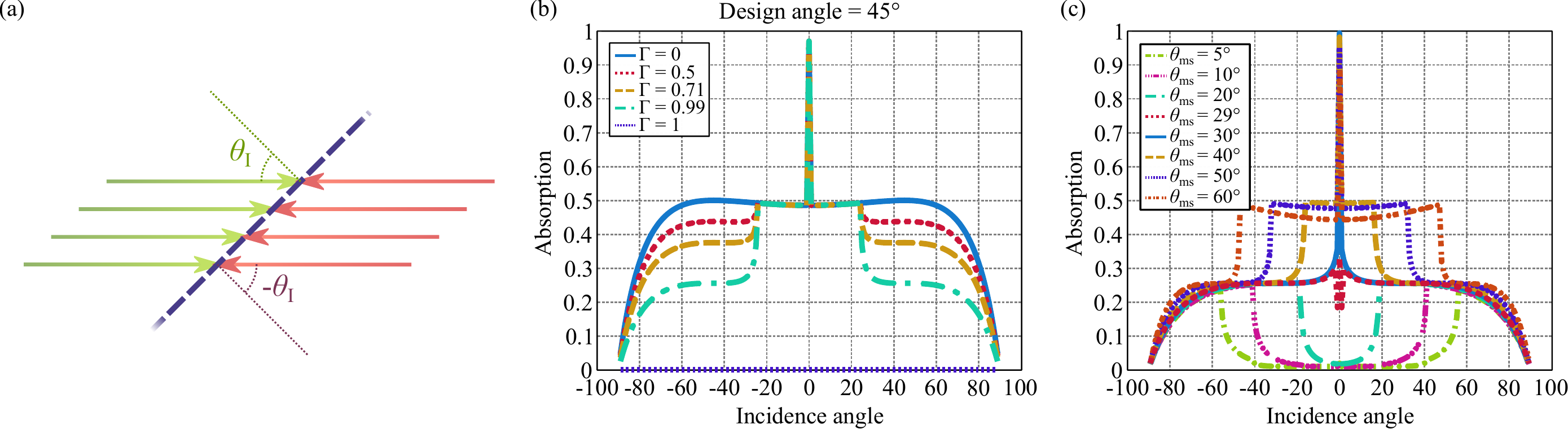}
\caption{Positioning a coherent asymmetric absorber in the field of two counter-propagating plane waves allows us to obtain a structure capable to sense small changes in the metasurface angular position. Coherent illumination at the same angle is realized only for the normal incidence ($\theta\@{I}=0$), in which case the absorption it total. Absorption for any non-normal illumination is comparable to the single-source scenario, where the absorption is limited by $50\%$. (a) Geometry of the setup.   
(b)--(c) 
The angular dependence of absorption at oblique illumination can be controlled tuning $\theta\@{ms}$ and $\Gamma$. The example in (b) is for the design angle $\theta\@{ms}=45^{\circ}$ and varying $\Gamma$, while (c) is for the reflection coefficient of $\Gamma=0.99$ and variable design angle.} \label{fig:angular_sensing}
\end{figure*}

The second application exploits the fact that for any mismatched illumination the absorption is limited by $50\%$ of the incident power, while it can be perfect at coherent and symmetric illumination.  The sources (or one source and a uniform mirror) are aligned so that the two incident waves propagate in the opposite directions, forming a standing wave. The metasurface is positioned in the field of this standing wave with a tilt with respect to the propagation axis, as illustrated in Fig.~\ref{fig:angular_sensing}(a). From the point of view of the metasurface, one of the sources illuminates it at a positive angle of incidence $\theta\@{I}$, while the other wave has a negative angle of incidence $-\theta\@{I}$. Thus, the condition of coherent illumination that requires equal incident angle for both waves, is not fulfilled. The only exception, when the metasurface receives matched illumination, is the case of the normal incidence $\theta\@{I}=0$. At this singular value of the tilt angle, the absorption is perfect, as shown in Figs.~\ref{fig:angular_sensing}(b)--(c). For any other angle of incidence, the absorption level drops to $50\%$ or more [see Fig.~\ref{fig:scattering_results}]. The profile, shown in  Fig.~\ref{fig:angular_sensing}(b), is  symmetric, as either both waves illuminate in the absorption regime, or one of them excites the non-specular mode.  Absorption level for angles compatible with non-specular reflection can be controlled using the parameter $\Gamma$, down to $25\%$ of the total incident power. The angular range of these  regions of different absorption  can be controlled through the design angle $\theta\@{ms}$, as shown in Fig.~\ref{fig:angular_sensing}(c). The  case when the metasurface is designed for $\theta\@{ms}=30^{\circ}$ is of special attention, as the non-specular mode can only be excited as a propagating wave for positive angles of incidence $\theta\@{I}$. As a result, considering a reflection coefficient value of $\Gamma=0.99$, the difference between the absorption peak at normal illumination and the absorption at oblique illumination is maximized.

\section{Conclusions}

In this paper we presented the concept of coherent asymmetric absorber that changes its behavior according to the incidence angle of  coherent-wave illuminations. It was shown that negligibly thin nonuniform sheets characterized by the determined sheet impedance operate as coherent counterparts of asymmetric absorbers. Under coherent illumination by two plane waves, absorption changes from nearly perfect to very small when the incidence angle varies. The study of arbitrary illuminations of these sheets reveals possibilities to tune the absorption and reflection changing the amplitude and phase of one of the two illuminating waves. 

Furthermore, this paper proposed two applications of coherent asymmetric absorbers: as broadband asymmetric waveguide absorbers and sensors of  angular position. By placing a coherent asymmetric absorber sheet inside a parallel-plate waveguide, equidistant to its walls, it is possible to realize broadband absorption without a need to block the cross-section opening of the waveguide. 

Exploiting the dramatic difference between achievable absorption under perfect coherent illumination by two waves with the same incident angle and any other illuminations, we showed that it is possible to detect angular displacements of objects with extreme accuracy.

\bibliographystyle{IEEEtran}

\bibliography{References_absorbers}

\end{document}